\documentclass[floatfix,aps,twocolumn,prd,showpacs,amsmath,nofootinbib,preprintnumbers]
{revtex4}

\newcommand{\eql}[1]{\label{eq:#1}}
\newcommand{\be}{\begin{equation}}
\newcommand{\ee}{\end{equation}}
\newcommand{\ec}[1]{Eq.~(\ref{eq:#1})}
\newcommand{\rf}[1]{\ref{fig:#1}}
\newcommand{\bpf}{{\rm BPF}}
\def\bea{\begin{eqnarray}}
\def\eea{\end{eqnarray}}
\usepackage{graphicx}
\usepackage{multirow}

\begin{document}

\title{Backgrounds and Projected Limits from Dark Matter Direct Detection Experiments}

\author{Scott Dodelson}
\email{dodelson@fnal.gov}
\affiliation{Center for Particle Astrophysics, Fermi National Accelerator Laboratory,
Batavia, IL~~60510-0500}
\affiliation{Department of Astronomy and Astrophysics, University of Chicago, Chicago, IL 60637, USA}
\affiliation{Kavli Institute for Cosmological Physics, Chicago, IL 60637, USA}

\date{\today}

\begin{abstract}
A simple formula is introduced which indicates the amount by which projections of
dark matter direct detection experiments are expected to be degraded due to backgrounds. 
\end{abstract}

\maketitle

\section{Introduction}

Experiments sensitive to weakly interacting dark matter particles are placing ever tighter constraints on 
dark matter-nucleon cross sections (e.g., \cite{cdms,xenon,warp}). The most recent upper limits rule out a substantial chunk of parameter
space for well-motivated theories (see, e.g., \cite{Jungman:1995df}). Even more exciting are the plans for future experiments which have excellent discovery
potential~\cite{dmsag}. 

As these experiments get larger and more expensive, it becomes important for the community to agree on uniform standards for
projecting the sensitivity of future experiments. Projections are often carried out simply by scaling exposures, defined as the product of fiducial detector mass $M$ times integrated detector live time $T$. That is, the projected improvement in the upper limit
on the cross section is taken to be
\be
\frac{\sigma^{\rm UL}}{\sigma^{\rm UL,curr}} \rightarrow \frac{1}{\lambda}  \equiv \frac{(MT)^{\rm curr}}{MT}
.\ee

Neglected in these projections is the effect of backgrounds. The purpose of this note is to provide a simple 
formalism that
accounts for the degradation of the future experiment's reach due to
backgrounds. A given experiment's projected constraint on the cross
section should be multiplied by the {\it background penalty factor} introduced here to get a baseline estimate of that
experiment's constraining power. This baseline estimate assumes that the future incarnation has the same background rate per
volume as the current version of the experiment. Experimentalists could plausibly argue that they expect to reduce their background
rates by some factor (here defined as $r$) and therefore they project a tighter constraint than the baseline estimate. 
The framework introduced here can also incorporate these arguments by computing the background penalty factor as a function of $r$. 

\section{Likelihood for Direct Detection}

Consider a direct detection experiment with fiducial detector mass $M$ operating
for a time $T$. What is the likelihood that such an experiment will observe $N$
events? We will assume that there are two sources of events: signal and
background. The expected number of signal events is 
\be
N_s = a\sigma M T \eql{nsdef}
\ee
where $\sigma$ is the dark matter-nucleon cross section we are trying to
constrain and $a$ is some
known constant\footnote{Different incarnations of the same experiment may have different 
fiducial regions, thereby changing the effective value of $a$. We choose to incorporate this effect into the definition
of $M$, which will be the mass of the fiducial region of the detector.}. The expected number of background events is
\be
N_b = r N_b^{\rm curr} \lambda
\ee
where $N_b^{\rm curr}$ denotes the number of background events expected in the most
recent data set of the experiment
and $r$ is the expected background rate normalized to its current value. 
If $r=1$, the future
version of this experiment will have the same background rate per volume as the
current one. If $r<1$ then the experiment projects to do better at background
rejection than it is currently doing (even though there may be new sources of
background at lower rates showing up in the new version).

The probability of seeing $N$ events given $\sigma$ is:
\be
\mathcal{L}(N\vert N_s(\sigma)) = \frac{(N_s+N_b)^N e^{-[N_s+N_b]}}{N!}
.\label{eq:like}\ee
The Bayesian 95\% CL upper limit on the cross section, $\sigma_{\rm UL}$, satisfies
\be
\int_0^{N_s(\sigma_{\rm UL})} dN_s \mathcal{L}(N\vert N_s) = 0.95 \int_0^\infty dN_s \mathcal{L}(N\vert N_s)
.\eql{ul}\ee
We want to project the constraints from future experiments so do not know the observed value $N$. In principle,
projecting future constraints requires the following steps:
\begin{itemize}
\item Generate a number of events $N$ from the distribution in \ec{like}
\item Compute the likelihood in \ec{like} using this value of $N$ as a function of $N_s$
\item Use \ec{ul} to determine the 95\% upper limit on $N_s$ for this realization
\item Repeat a large number of times and average all the upper limits
\end{itemize}
In practice, we have found that this is not necessary, that the averaged projected upper limit is within a few
percent of that obtained by simply setting $N=N_b$ (we assume no signal and project upper limits). 
\ec{ul} determines $N_s(\sigma_{\rm UL})$ as a function of $N_b$. The lower curve in Fig.~\ref{fig:nbnsint} shows this
limit on $N_s$
as a function of the number of background events in the case we have been considering so far, when the
number of expected background events is known precisely. For example, an experiment which has no 
expected backgrounds is projected to rule out $N_s>3$. An experiment with one hundred expected background events
(and no uncertainty on this background prediction)
projects to rule out all cross section which predict $N_s>20$.

\newcommand{\sfig}[2]{
\includegraphics[width=#2]{#1}
		}
\newcommand{\Sfig}[2]{
	\begin{figure}[htbp]
	\sfig{#1.eps}{0.9\columnwidth}
	\caption{{\small #2}}
	\label{fig:#1}
	\end{figure}
}
\Sfig{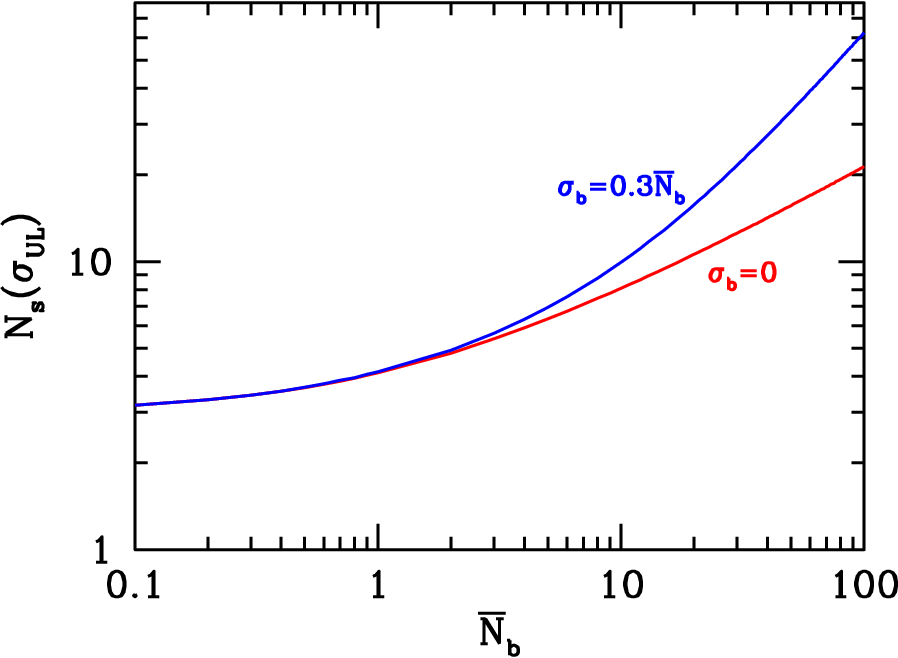}{Number of events expected from the signal at its maximally allowed
value as a function of the number of background events in the experiment. For
example, if there is one expected background event, then the expected upper
limit on the number of signal events is 4.1. This means that all values of
the cross-section which predict more than 4.1 events in the detector are ruled
out. Lower (red) curve shows this limit when there is no uncertainty in the number
of expected background events. Upper (blue) curve shows the effects of a 30\% uncertainty
in the number of background events.}

More generally there will be uncertainty in the expected backgrounds, so the probability of observing
$N$ events is actually an integral over the distribution of possible values of $N_b$. Assume that
$N_b$ is drawn from a Gaussian distribution with 
mean $\bar N_b$ and variance $\sigma_b^2$. Then, \ec{like} gets generalized to 
\bea
\mathcal{L}(N\vert \sigma) &\propto& \int_0^\infty dN_b\, 
\exp\left\{\frac{-(N_b-\bar N_b)^2}{2\sigma_b^2}
\right\} \nonumber\\
&&\times\frac{(N_s+N_b)^N e^{-[N_s+N_b]}}{N!}
.\eea
We can again determine $N_s(\sigma_{\rm UL})$ from this likelihood function using \ec{ul}. So for fixed $\bar N_b$
and $\sigma_b$,
one arrives at a projected upper limit on the number of signal events. This upper limit then is a function of $\bar N_b$ and $\sigma_b$. Let's call this function $F(\bar N_b,\sigma_b)$. It monotonically
increases as $\bar N_b$ increases and tends to 3 for small values of $\bar N_b$. For large values of $\bar N_b$, the uncertainty in the background adds in quadrature with the Poisson uncertainty, so the uncertainty on the number of events is $\sqrt{\bar N_b + \sigma_b^2}$. 
We know the expected number of events with this uncertainty so can rule out 
a model which produces $N_s(\sigma_{\rm UL}) \ge 2\sqrt{\bar N_b + \sigma_b^2}$ at the 2-sigma level. This indeed is the behavior of the $\sigma_b=0.3\bar N_b$ curve in Fig.~\rf{nbnsint}.

As indicated in Fig.~\rf{nbnsint}, in the absence of backgrounds an experiment can rule out models
which predict $N_s>3$. Given \ec{nsdef}, this translates into an upper limit on the cross section of
\be
\sigma_{\rm UL} = \frac{3}{aMT} \qquad \qquad ({\rm Zero\ Background})
.\ee
The ratio of the upper limit in a future experiment to the current one would then be:
\be
\frac{\sigma_{\rm UL}}{\sigma^{\rm curr}_{\rm UL}}
= \frac{1}{\lambda}.
\ee
This is the projection typically offered for a future experiment. Fig.~\rf{nbnsint} offers a way to 
flesh out this projection accounting for backgrounds.

\section{Background Penalty Factor}

Consider an experiment which currently expects 10 background events, so is able to rule
out all models which predict more than 10 events (from now on assuming $\sigma_b=0.3\bar N_b$ and reading off from
Fig.~\rf{nbnsint}).
If the background rate remains the same, then a future version of this experiment
with ten times the exposure will expect to see 100 background events. Reading off from Fig.~\rf{nbnsint} 
at $\bar N_b=100$, we see that $N_s(\sigma_{\rm UL})=62$. So, the projected upper limit from the future
version is
\be
a10MT\sigma_{\rm UL} = 62
\ee
Divide this by the old upper limit with $MT$ ten times smaller:
\be
aMT\sigma^{\rm curr}_{\rm UL} = 10
\ee
and we see that the upper limit on the cross section projects to be tighter by a factor of
$62/(10\times 10)=0.62$. So the limit gets better only by a factor of $1/.62=1.6$ 
with a factor of 10 increase in mass.

This suggests defining a {\it Background Penalty Factor} (BPF). In the absence of any backgrounds, the upper limit on
the cross section will go down by the amount by which the exposure increases. In fact, though, this improvement will be attained only if the background
rates are reduced considerably. The BPF is the
{\it ratio of the projected limit assuming a background rate $r$ and 
the limit assuming zero backgrounds}.  
The equation which defines it is
\be
\bpf \equiv \lambda \frac{\sigma_{\rm UL}}{\sigma_{\rm UL}^{\rm curr}}
= \frac{F(r\lambda N_b^{\rm curr})}{F(N_b^{\rm curr})}
\ee
where again $F$ is the function plotted in Fig.~\rf{nbnsint}.
The current practice is to set BPF=1, so that the expected upper limit scales simply as $1/\lambda$. The actual
scaling is $BPF/\lambda$.  
In the example
above, $\bpf = 62/10$. Fig.~\rf{bpfgrid1} shows the BPF as a function of the number of current background
events and the product $r\lambda$ with $\sigma_b\rightarrow 0.3\bar N_b$. 

\Sfig{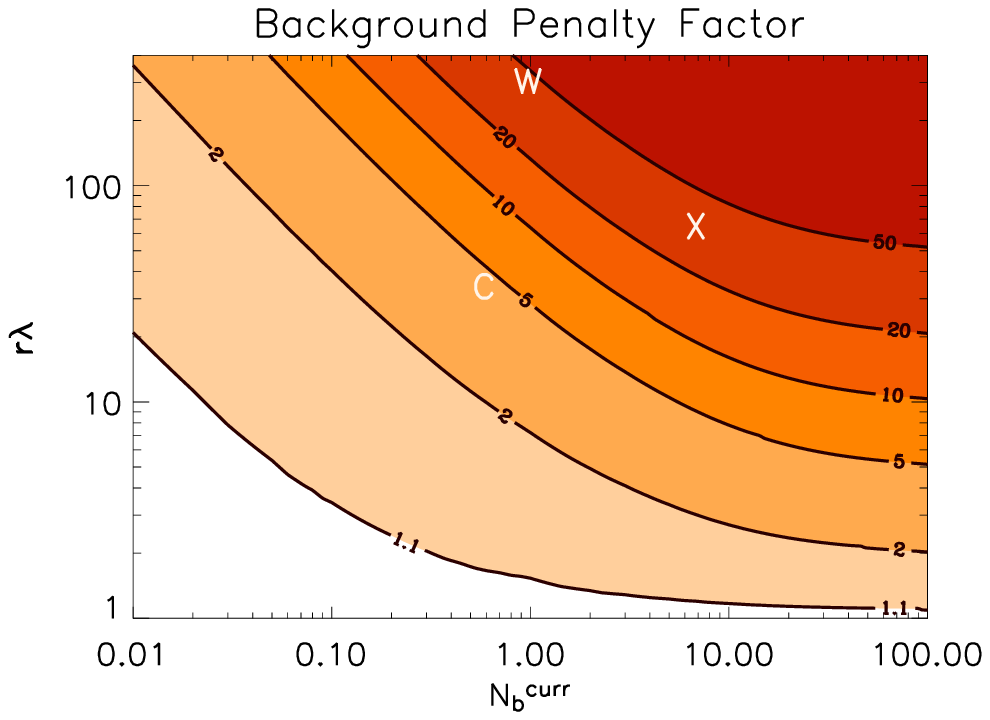}{Background Penalty Factor as a function of number of background events in the current incarnation of the
experiment and the product $r\lambda$ where $r$ is the expected background rate in the future experiment compared with the current
rate and $\lambda$ is the ratio of exposure ($M\times T$) in the future vs. current experiments. Here the uncertainty of the
predicted background is set to $\sigma_b=0.3\bar N_b$. Three experiments are shown with
$r$ set to one: C is for SuperCDMS; X for Xe100; and W for WARP140.}

It is important to underscore the uncertainties in these projections. While ignoring backgrounds (as do some of the current projections) 
does not seem appropriate, neither can we assume that the background rates will remain constant. Better shielding, deeper sites, and purified
samples are among the tools groups are using to reduce their background rates. A fairer statement is that the projected sensitivity
of a future experiment is bracketed by the usual zero background (ZB) projection and the projection with background rates set equal to
their currently achieved values ($r=1$). 

With this in mind, the BPF's of three different future experiments are depicted in Fig.~\rf{bpfgrid1} with $r$ set to one: 
the 25 kg version of Super CDMS, Xenon 100 (LUX would be similar), and WARP 140. The goal here is not to be comprehensive but rather to apply this formalism to a handful of the leading experiments. Further, the parameters chosen here for these experiments are uncertain and have not been vetted by the collaborations.
In each case, the current level of backgrounds comes from their most recent result (0.6 for CDMS, 
7 for Xenon10, and 1 for WARP). Table I lists the zero-background projection and the current-background-rate (CBR) projection for these three future experiments.

\Sfig{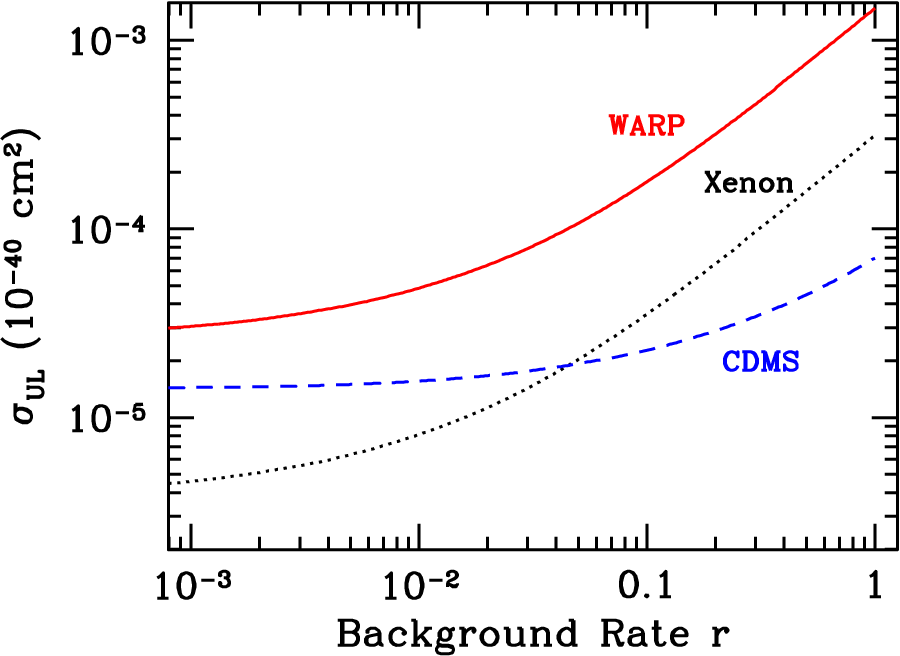}{Projected upper limits on dark matter cross sections for three different experiments -- Super CDMS (blue),
WARP 140 (red), and Xe100 (black) -- as a function of background rate for $m=60$ GeV. A future rate per volume equal to their most recent experiment
corresponds to $r=1$.}

Fig.~\rf{rlimit1_signb} shows
how much better each experiment can do as a function of how well 
they reduce backgrounds. If SuperCDMS reduces its background rate by a factor of
ten, then its limit on the cross section would be $3\times 10^{-45}$ cm$^2$, a factor of two within its ultimate reach.
Xenon100 and LUX have more potential reach: if their background
rates can be cut by a factor of a hundred, they can reach cross sections below $10^{-45}$ cm$^2$
for a dark matter mass of 60 GeV. WARP too can be competitive with SuperCDMS if its background rate can be
reduced by a factor of a hundred.

\section{Conclusions}

Projections of the future reach of dark matter experiments should account for backgrounds. The Background Penalty
Factor introduced here is a way to go beyond the standard projection which assumes
zero backgrounds. This penalty factor is a function of the expected background rate, so 
projections using BPF, as illustrated in Fig.~\rf{rlimit1_signb}, can quantitatively inform 
experimenters (and reviewers) the extent to which background rates need to be reduced.

{\bf Acknowledgments:}
This work was supported by the Fermi Research Alliance, LLC under Contract No. DE-AC02-07CH11359 with the US Department of Energy. I thank Dan Bauer, Richard Gaitskell, Bernard Sadoulet, Richard Schnee, and Steve Yellin for helpful comments and discussions. Fortran code to compute the BPF is available at {\tt http://home.fnal.gov/\~{}dodelson/dm.html}.

\begin{widetext}
 
\begin{table}[h]
\caption{Range of Projections for three future dark matter direct detection experiments. Background Penalty Factor
with current background rate is given in the fourth column. Fifth column lists the current upper limit on
the cross section of a dark matter particle with $m=60$ GeV. Sixth column is the zero background projection of
the upper limit in this experiment obtained by scaling the current limit by the ratios of exposures (2nd and 3rd columns).
Last column degrades this limit by the BPF assuming Current Background Rates ($r=1$). 
The actual limit will likely lie in the range delineated by these last two columns.
}
\begin{tabular}{| c || c |c| c| c |c| c| }
\hline
\multirow{2}{*}{Experiment} & Current Exposure  & Future Exposure 
& {BPF} & Current Limit & ZB Projection & CBR Projection  \\
& (kg-yrs) & (kg-yrs) & $(r=1)$  & (cm$^2$) & (cm$^2$)& (cm$^2$)\\
\hline
\hline
CDMS & 0.33 & 10 &   3.9 & $6.6\times 10^{-44}$ & $2.2\times 10^{-45}$ & $8.6\times 10^{-45}$\\
Xenon & 0.44 & 25  & 29.5 &  $6\times 10^{-44}$& $1.1\times 10^{-45}$ & $3.2\times 10^{-44}$\\
WARP & 0.13 & 35 &  39.7  & $10^{-42}$ & $3.7\times 10^{-45}$ &  $1.5\times 10^{-43}$ \\
\hline
\end{tabular}
\end{table}

\end{widetext}

\end{document}